\documentclass[reprint,prl,twocolumn,twoside,aps]{revtex4-1}

\usepackage{graphicx}

\usepackage{amsmath, amssymb}
\usepackage{bm}
\usepackage{natbib}
\usepackage{color}
\usepackage{soul,xcolor}

\newcommand{\kew}[1]{{\color{red} #1}}
\setstcolor{red}

\begin{document}

\title{Observation of photon droplets and their dynamics} 
\author{Kali  E. Wilson$^{1,*}$, Niclas Westerberg$^{1}$, Manuel Valiente$^1$, Callum W. Duncan$^1$, Ewan M. Wright$^{2,1}$, Patrik \"Ohberg$^1$, Daniele Faccio$^{1,3}$}
\email{kali.e.wilson@durham.ac.uk, Daniele.Faccio@glasgow.ac.uk\\}
\affiliation{$^1$Institute of Photonics and Quantum Sciences, Heriot-Watt University, Edinburgh EH14 4AS, UK} 
\affiliation{$^2$College of Optical Sciences, University of Arizona, Tucson, AZ, 85721, USA}
\affiliation{$^3$School of Physics \& Astronomy, University of Glasgow, Glasgow G12 8QQ, UK}


\begin{abstract}
We present experimental evidence of photon droplets in an attractive (focusing) nonlocal nonlinear medium.  Photon droplets are self-bound, finite-sized states of light that are robust to size and shape perturbations due to a balance of competing attractive and repulsive forces.  It has recently been shown theoretically, via a multipole expansion of the nonlocal nonlinearity, that the self-bound state arises due to competition between the $s$-wave and $d$-wave nonlinear terms, together with diffraction. The theoretical photon droplet framework encompasses both a soliton-like stationary ground state and the non-soliton-like dynamics that ensue when the system is displaced from equilibrium, i.e. driven into an excited state.  We present numerics and experiments supporting the existence of these photon droplet states and measurements of the dynamical evolution of the photon droplet orbital angular momentum.  
\end{abstract}

\maketitle

\begin{figure}[t!]
\centering
\includegraphics[width=0.95\columnwidth]{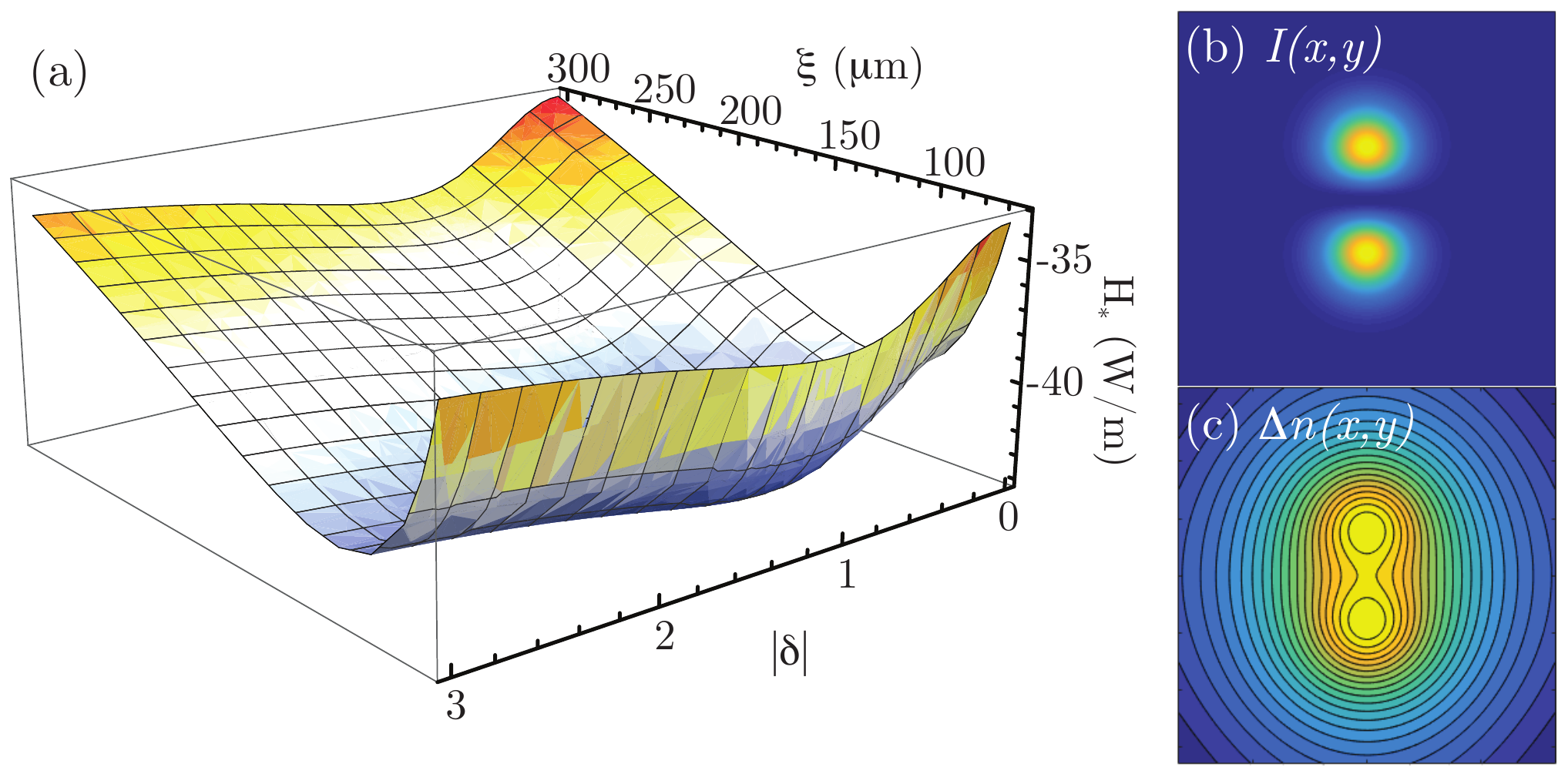}
 \caption{(a) Pseudo-energy landscape $H_*(\delta,\xi)$ for $P_0 = 0.8$ W and $E_0(r) \propto r e^{-r^2/\xi_0^2}$.  (b) and (c) $320~\mu$m x $320~\mu$m square images showing the transverse intensity profile $I(x,y)$ and nonlinear potential $\Delta n(x,y)$, respectively, for the ground state.  The minimum pseudo-energy occurs for $|\delta| = 1$ and $\xi \sim 100~\mu$m, and there is an instability at $H_*(\delta = 0)$ in contrast to the SM model. }
 \label{fig:landscape}
 \end{figure}

Vortex beams in local nonlinear media experience symmetry breaking azimuthal instabilities \cite{ Chiao1964, Desyatnikov2005, Tikhonenko1995, Petrov1998}.  Nonlocal nonlinearities can stabilize such beams \cite{Desyatnikov2012, Suter1993}; solitary-wave behavior for both Laguerre-Gauss (LG) \cite{Rotschild2005} and Hermite-Gauss (HG) modes \cite{Rotschild2006} has been demonstrated experimentally with a laser beam propagating through a thermal nonlinear medium.  The theoretical treatment for such behavior stems from the Snyder-Mitchell (SM) model for accessible solitons \cite{Snyder1997} where the effective potential due to the attractive (focusing) nonlocal nonlinearity is approximated by a parabolic function, with a local minimum at $r~=~0$ \cite{Lopez-Aguayo2006, Buccoliero2007, Buccoliero2008}.  For sufficiently high nonlocality, a vortex beam with orbital angular momentum (OAM) constitutes a stable soliton solution of the nonlocal nonlinear Schr\"odinger equation (nonlocal NLSE) \cite{Yakimenko2005}.  However, the nonlocality must be much greater than the beam size in order to allow the effective nonlinear potential to remain parabolic regardless of the shape of the input beam \cite{Westerberg2017, Petrovic2017}.  The dynamics of structured beams, such as dipole solitons \cite{Shih1997, Fratalocchi2007}, azimuthons \cite{Minovich2009, Lopez-Aguayo2006} and higher-order HG beams \cite{Izdebskaya2013}, have been explored either within the SM potential or through a purely numerical treatment of the nonlocal nonlinearity.\\
\indent Recent theoretical work has introduced the concept of photon droplets, or droplets of light, defining them as ``self-bound, finite-sized objects that are stable against perturbations in size, shape and density due to a competition of attractive and repulsive forces'' \cite{Westerberg2017}.  We consider a soliton to be a beam whose transverse spatial profile does not change under propagation. The definition of a photon droplet does not require stationarity, however, if a droplet is initiated in the ground state, then its stationary behavior will coincide with the solitons described above. Liquid light states have also been proposed, arising from the balance between higher order, cubic-quintic nonlinearities \cite{Michinel2006, Alexandrescu2009, Adhikari2016} that are distinct from the photon droplets.  In the photon droplet framework, a beam of light propagating in a nonlocal nonlinear medium is treated as a many-body system where the photon-photon interaction is mediated by the nonlinearity $\Delta n$.  A multipole expansion of $\Delta n$ allows calculation of a pseudo-energy landscape for the many-body photon state \cite{Westerberg2017}, as a function of the state's physical size and net OAM.  The expansion reveals competition between an $s$-wave nonlinear term favoring a stable vortex ring \cite{Yakimenko2005} and a $d$-wave nonlinear term favoring a two-lobed structure (see Fig.~\ref{fig:landscape}(b)). This competition together with kinetic energy (diffraction) results in a robust, $p$-wave-symmetric self-bound state, with a lower pseudo-energy than that of the azimuthally symmetric SM vortex soliton \cite{Westerberg2017}. The energy-landscape description enabled by the multipole expansion of $\Delta n$ encompasses both a stationary lowest energy state with $p$-wave symmetry and zero net OAM, and the dynamics of a broader class of excited states including the rotating dipole solitons and azimuthons discussed above. Photon droplets extend beyond these concepts, attain liquid-like properties, and exhibit a liquid-to-gas transition concurrent with the emergence of sound waves as low energy excitations. Furthermore their pseudo-energy per unit power, analogous to the energy per particle \footnote{As this is an out-of-equilibrium system, an EoS cannot be defined and the energy per particle is the relevant quantity with which to compare.}, has the same form as the equation of state (EoS) for droplets of one-dimensional liquid Helium \cite{Westerberg2017, Valiente2016} and shares similarities with the EoS for droplets found in Bose-Einstein condensates (BECs) \cite{FerrierBarbut2016, Schmitt2016, Chomaz2016, Cabrera2017, Cheiney2017}.  Whilst inherently quantum, BEC droplets are often described by a classical Gross-Pitaevskii equation, where an effective classical potential accounts for the quantum-fluctuation stabilization term \cite{Schmitt2016}. \\
\indent In this Letter we provide the first experimental evidence of photon droplets and study their evolution. We focus attention on the out-of-equilibrium dynamics characterized by evolution of the droplet angular momentum.  Experimental OAM-mode decomposition measurements show that the net OAM of the photon droplet varies as it propagates.  The droplet and higher-order OAM excitations exchange OAM back and forth as the photon droplet explores its pseudo-energy landscape in a manner reminiscent of the two-way energy exchange between a hot spot and the photon bath in optical filamentation \cite{Couairon2007}. 

\indent In our experiment, a beam with a $p$-wave symmetric intensity profile propagates through an attractive (focusing) nonlocal nonlinear medium, in this case lead-doped glass (SF6) with a thermal nonlinearity characterized by $\beta = -dn/dT = 14 \times 10^{-6} ~\mathrm{K}^{-1}$, a heat capacity of $\kappa = 0.7~ \mathrm{W} \mathrm{m}^{-1} \mathrm{K}^{-1}$, linear absorption $\alpha = 0.01~\mathrm{cm}^{-1}$, and background index of refraction $n_0 = 1.8$ \cite{Roger2016}.   The propagation of the electric field $E(\mathbf{r},z)$ through this medium is described by the NLSE
\begin{equation}
i\frac{\partial E}{\partial z} = -\frac{1}{2k_0}\nabla^2 E-\frac{k_0}{n_0}\Delta n E -\frac{i\alpha}{2}E\equiv \mathcal{H}_*E.  \label{paraxial}
\end{equation}
Here $k_0 = 2\pi n_0 / \lambda$ is the wavenumber in the medium, for light of vacuum wavelength $\lambda$, $\nabla^2$ is the two-dimensional Laplacian for the transverse plane $\mathbf{r} = (x, y)$.   In analogy to BECs, $\mathcal{H}_*$ is the pseudo-energy density, $E$ plays the role of the many-body wavefunction, and the propagation direction $z$ maps to time \cite{Pethick2008}. The nonlocal nonlinearity $\Delta n = \gamma \int d^2r' R(\mathbf{r}-\mathbf{r}')|E(\mathbf{r}',z)|^2$, is well described by the real space response function $R(\mathbf{r}) = K_0(|\mathbf{r}|/\sigma)/2\pi \sigma^2$.
Here $\gamma = \alpha \beta \sigma^2 / \kappa = 1.25 \times 10^{-6} ~\mathrm{cm}^2~\mathrm{W}^{-1}$ gives the nonlinear coefficient characterized by the physical system parameters, and $K_0$ is the zeroth-order modified Bessel function of the second kind. The nonlocal length $\sigma = D/2 = 2.5$ mm is fixed to half the smallest dimension of the material, which is a valid assumption for a rectangular geometry in the steady-state regime with $\Delta T = 0$ at the boundary \cite{Vocke2016, Roger2016}.  Over the extent of the beam the measured response function is to a good approximation radially symmetric.
We evaluate the pseudo-energy functional \cite{Westerberg2017}
\begin{equation}
H_*=\int d^2r' E^*(\mathbf{r}',z)\mathcal{H}_*(\mathbf{r}',z)E(\mathbf{r}',z),\label{pseudoenergy}
\end{equation}
where $\mathcal{H}_*$ is defined in Eq.~(\ref{paraxial}), and the corresponding Lagrangian density is $\mathcal{L}=E^*\left(i\partial_z - \mathcal{H}_*\right)E$.
To evaluate $H_*$ analytically, we choose the ansatz 
\begin{equation}
E(\mathbf{r},0)\propto r e^{-r^2/\xi_0^2}\left[e^{i\phi}+\delta_0 e^{-i\phi}\right] \label{ansatz}
\end{equation}
for the input beam, where $\phi$ is the azimuthal angle in cylindrical coordinates, $\xi_0 = \xi(z = 0)$ is the $1/e^2$-intensity beam radius, and $\delta_0 = \delta(z = 0)$ gives the ratio of OAM $\ell = -1$ to $\ell = +1$ \footnote{Note that this ansatz is of slightly different functional form than the droplet found in Ref.~\cite{Westerberg2017}, i.e. it is Gaussian instead of exponential. The difference is however small, and the ansatz used in here corresponds closely to the experimental input and should be thought of as an initial condition.}.  We perform a multipole expansion of the nonlocal nonlinearity
\begin{equation}
\Delta n \approx \gamma\left[P-d_{\nu}\partial^{\nu}_{\mathbf{r}}+\frac{1}{2}Q_{\nu,\mu}\partial^{\nu}_{\mathbf{r}}\partial^{\mu}_{\mathbf{r}}\right]R(\mathbf{r}),\label{LWA}
\end{equation}
where $P$ is the power, $d_{\nu}=\langle x_{\nu}\rangle$ is the Cartesian ${\nu}$-component of the dipole moment, and $Q_{\nu,\mu}=\langle x_{\nu}x_{\mu} \rangle$ is the corresponding element of the quadrupole moment tensor.  We refer to this expansion of the nonlocal nonlinearity as the long wavelength approximation (LWA), which is valid for a tightly bound scenario ($\sigma >> \xi$) \cite{Westerberg2017}.  
For an input beam of the form of Eq.~\ref{ansatz}, the dipole moment cancels leaving a competition between the linear $s$-wave and quadrupole $d$-wave terms resulting in the pseudo-energy landscape  $H_*(\delta, \xi)$ shown in Fig.~\ref{fig:landscape}(a). In analogy with BEC droplets \cite{Schmitt2016, FerrierBarbut2016}, the bound state depends on the number of particles in the system, which in this case is the beam power.   As shown in Fig.~\ref{fig:landscape}, we find a clear minimum associated with $|\delta| = 1$, corresponding to a two-lobed photon droplet with zero OAM, and a power-dependent value of $\xi = \xi_\mathrm{min}$.  At higher powers, the depth of the pseudo-energy well increases, and $\xi_\mathrm{min}$ decreases \cite{Westerberg2017}.   Figures~\ref{fig:landscape}(b) and~\ref{fig:landscape}(c) show the transverse intensity profile $I(x,y)$ and corresponding nonlinear potential $\Delta n(x,y)$ for the lowest-energy photon droplet.  
\begin{figure}[t!]
\includegraphics[width=0.95\columnwidth]{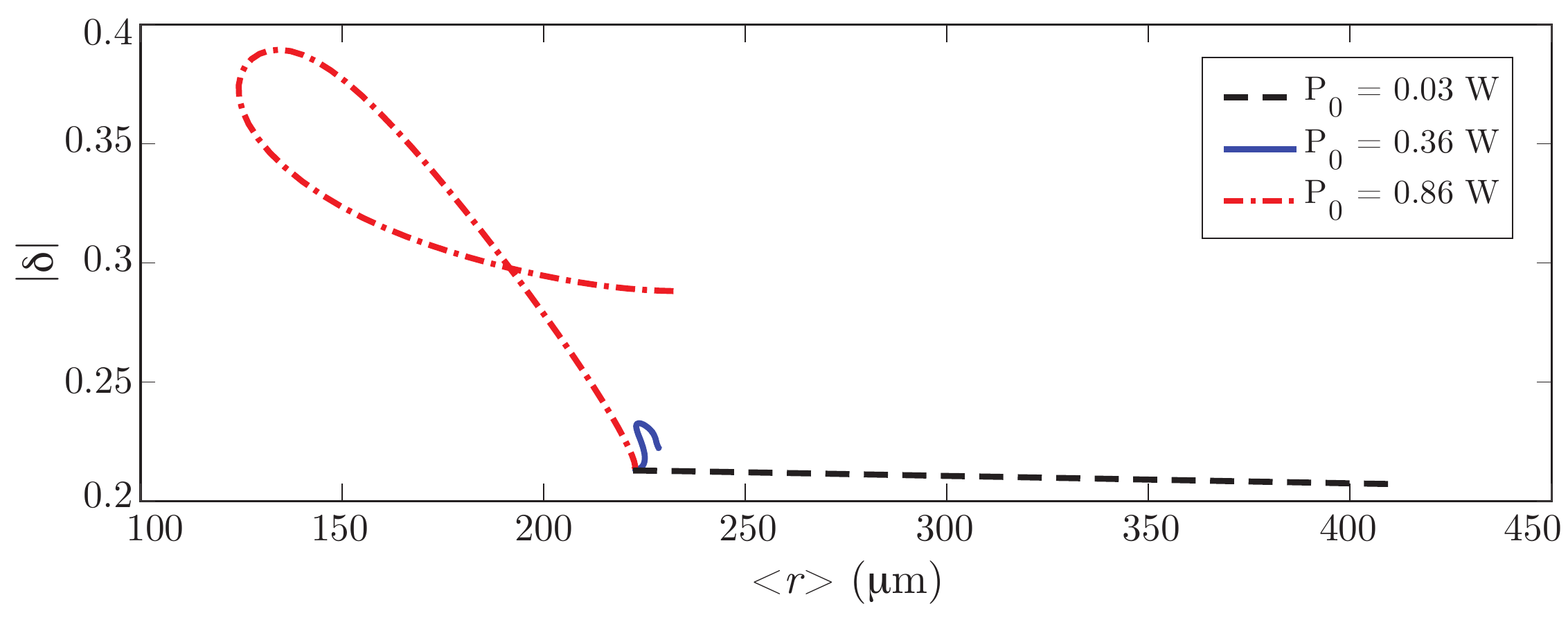}
   \caption{Photon droplet dynamics from numerics. Plot of $|\delta|$ versus $\langle r \rangle \propto \xi$ over 80 cm of propagation for $|\delta_0| = 0.2$, $\xi_0 = 210~\mu$m ($\langle r \rangle_0 = 220~\mu$m), and varying input beam powers $P_0$.  See supplemental materials for details of the spatial evolution.}
     \label{fig:OAM}
\end{figure}

\indent We solve Eq. \ref{paraxial} numerically \cite{Roger2016}, using the $k$-space response function $\hat{R}(k_x, k_y) = 1/(1 + \sigma^2(k_x^2 + k_y^2))$ \cite{Vocke2016} to account for the nonlocality associated with the nonlinearity, $\Delta \hat{n}(k_x,k_y)= \gamma \hat{R}(k_x,k_y) \hat{I}(k_x,k_y)$, where $\hat{I}(k_x,k_y)$ is the Fourier transform of  $I(x,y)$.  We map the photon droplet dynamics to the pseudo-energy landscape by calculating $|\delta| = |c_{-}|/|c_{+}|$ and the photon droplet radius $\langle r \rangle = \int r |E|^2 \; r dr d\phi ~/ \int |E|^2 \; r dr d\phi \propto \xi$ as the beam propagates.  Here $c_{+}$ ($c_{-}$) is the amplitude of the electric field associated with OAM $\ell = + 1$ ($\ell = -1$) calculated by decomposing the transverse electric field from the numerics onto an orthonormal basis of LG beams.  
\begin{figure}[t!]
\includegraphics[width=0.95\columnwidth]{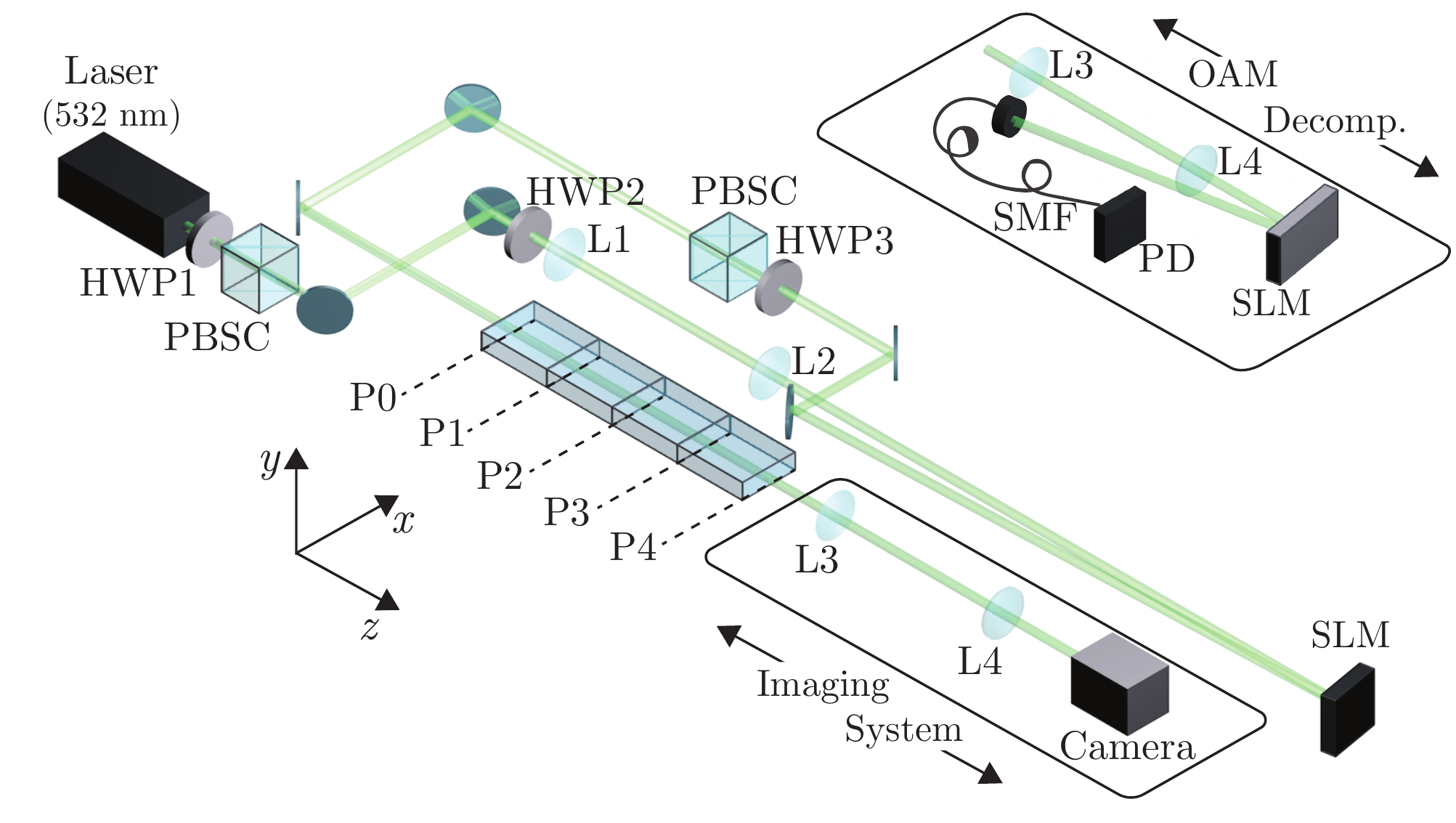}
 \caption{Experimental setup. Lenses L1 and L2 expand the 532-nm laser beam.  An SLM creates a diffraction grating with variable combinations of OAM $\ell = +1$ and $\ell = -1$. The first-order diffraction is weakly focused ($f \sim 2$ m) onto the entrance facet (P0) of a set of four  10-cm x 4-cm x 0.5-cm glass slabs.  HWP2 is used to vary the SLM diffraction efficiency.  HWP3 and PBSC provide additional control of $P_0$. The beam intensity profile is imaged at the entrance facet P0 and at the exit facet of each glass plate P1-P4.  The $M = 6$ imaging system is fixed with respect to the camera and translated as a single entity.  The camera can be replaced with an SLM and single-mode fiber to perform an OAM decomposition. }
  \label{fig:exp}
\end{figure}

\indent Figure~\ref{fig:OAM} shows plots of $|\delta|$ vs. $\langle r \rangle$ for varying input power over 80 cm of propagation, for initial conditions $|\delta_0| = 0.2$ and $\xi_0 = 210~\mu$m (${\langle r \rangle}_0 = 220~\mu$m). These plots show how the relevant physical parameters evolve as the photon droplet explores its pseudo-energy landscape $H_*(\delta, \xi \sim <r>)$. For low particle numbers (dashed black line, $P_0 = 0.03$ W) we observe near-constant $|\delta|$ as the radius increases, corresponding to linear diffraction in the absence of a bound-state or pseudo-energy well: the analogous particle number is less than what is required for a bound state to form, a common feature of droplets. For intermediate particle numbers (solid blue line, $P_0 = 0.36$ W) we observe small variations in both $|\delta|$ and $\langle r \rangle$, consistent with the low nonlinearity and correspondingly shallow pseudo-energy well at this power.  Here we have nearly matched the ground state radius $\xi_\mathrm{min} \sim 200~\mu$m for $P_0 = 0.36$ W, but $|\delta_0| = 0.2$ implies the photon droplet is displaced from equilibrium. Changes in $|\delta|$ must correspond to excitations of higher-order OAM modes such that the total  OAM of the system remains constant.   At $P_0 = 0.36$ W, there is a shallow pseudo-energy well (approximately 1/10 the depth of the one shown in Fig.~\ref{fig:landscape}(a)) but given that  the OAM of the system must be conserved, there is no path for the droplet to decay to lower energy; OAM conserving excitations are too costly.   For large particle numbers (dashed red line, $P_0 = 0.86$ W) we observe focusing in $\langle r \rangle$, and simultaneous variation in $|\delta|$.  The energy landscape is now steeper (see Fig.~\ref{fig:landscape}(a)) so high-energy excitations corresponding to higher-order OAM modes may be energetically allowed. \\
\indent Figure~\ref{fig:exp} shows our experimental setup. We use a 532-nm laser with a Gaussian transverse beam profile.  Lenses L1 and L2 expand the beam to a $1/e^2$-intensity beam radius of $\sim 3$ mm.   We use a spatial light modulator (SLM) to create a diffraction grating with variable combinations of OAM $\ell = +1$ and $\ell = -1$.  The SLM weakly focuses ($f \sim 2$ m) the first diffracted order onto the entrance facet P0 of a set of glass slabs (SF6).    The power in the beam is varied from $P_0 = 30$ mW corresponding to linear propagation to $P_0 = 0.86$ W. This results in an initial state $E(r,z = 0)$ at P0 that is a good approximation of the anzatz of Eq.~\ref{ansatz}.  The 40-cm long glass sample is composed of four slabs, each 10-cm x 4-cm x 0.5-cm, placed one after the other. We monitor the beam evolution at 10 cm intervals by imaging the output at successive planes, P0-P4, as indicated in Fig.~\ref{fig:exp}  {\footnote{\kew{The only relevant back-reflection (of order 8\%) is from the very first (P0) and last (P4) facets. The glass slabs are butted together with good optical contact between adjoining slabs, resulting in negligible back-reflections from any intermediate interfaces along the propagation direction.}}}. The camera and optics for the imaging system with magnification $M=6$ are shifted as a unit to image each plane in succession, allowing us to follow the evolution of the photon droplet in ``time" for a fixed input power. 
 \begin{figure}[t!]
\includegraphics[width=0.95\columnwidth]{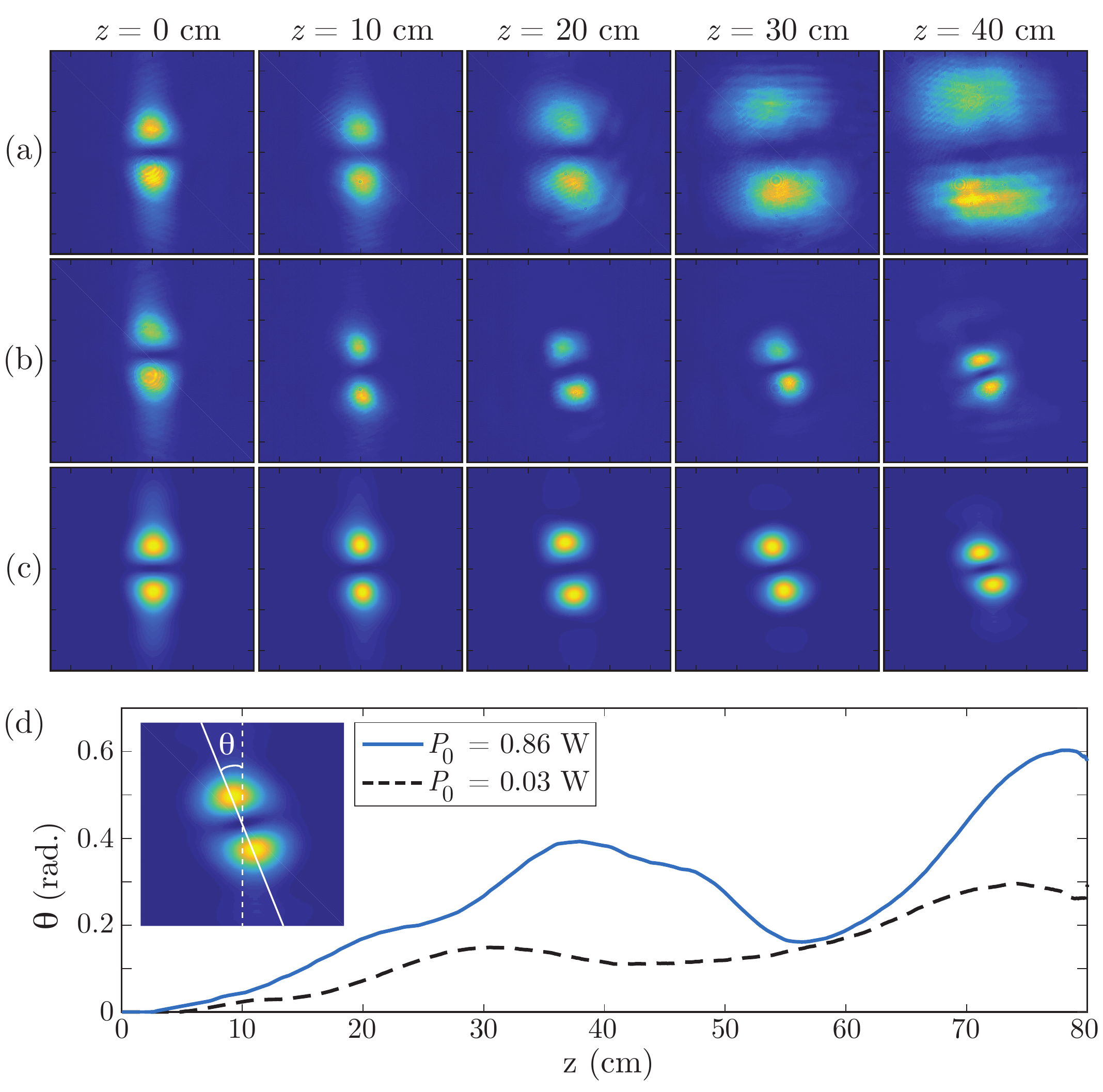}
   \caption{1-mm x 1-mm square images of the intensity profile for input conditions $|\delta_0| = 0.7$ and $\xi_0 = 160~\mu$m. (a) Experiment: effectively linear propagation for reference $P_0 = 0.03$ W. (b) Experiment: photon droplet propagation for $P_0 = 0.86$ W. (c) Numerics: propagation for $P_0 = 0.86$ W. (d) Numerics: plots of rotation angle $\theta$ versus $z$.}
     \label{fig:rotation}
\end{figure}
\begin{figure}[t!]
\includegraphics[width=0.95\columnwidth]{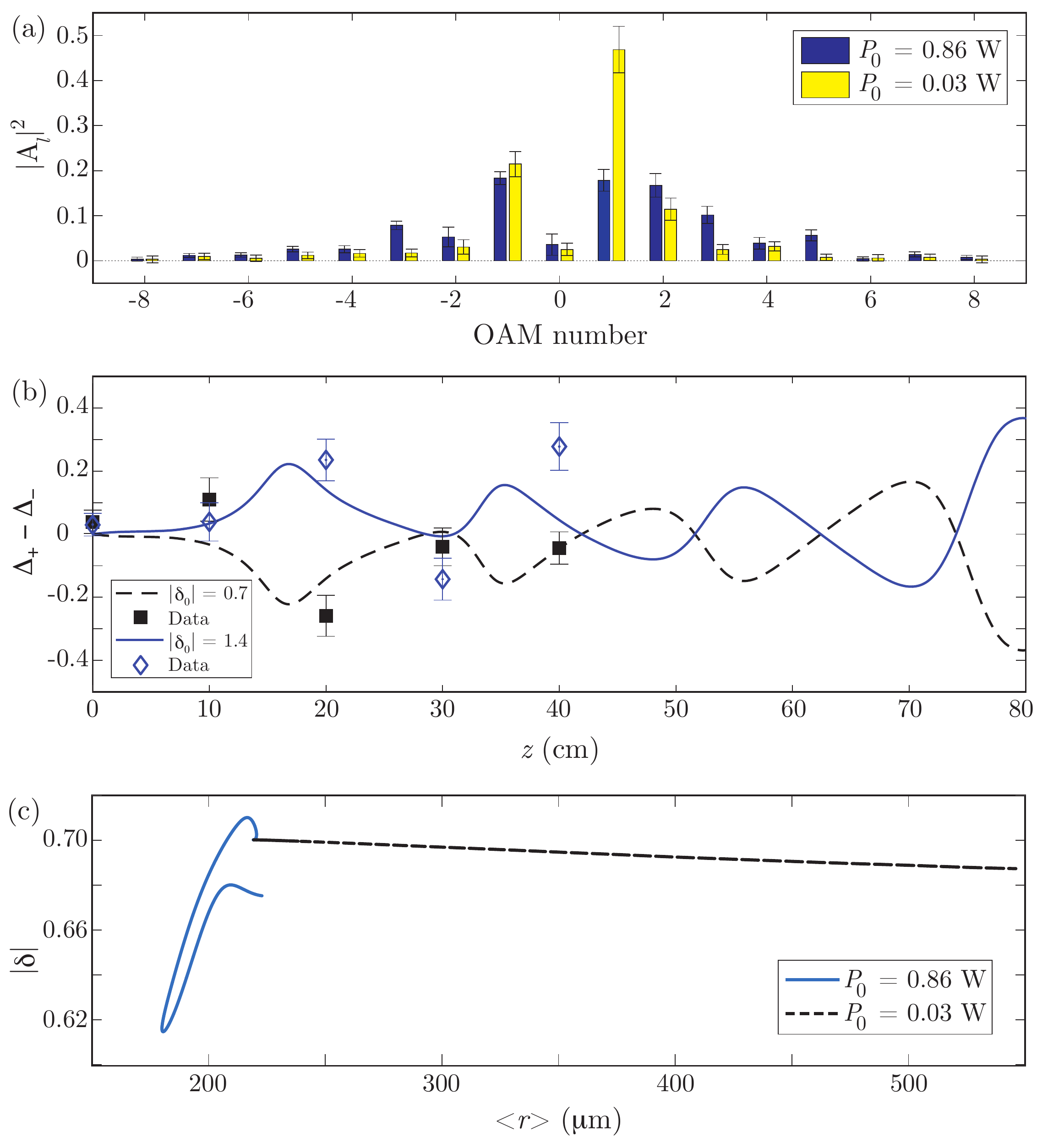}
   \caption{Photon droplet dynamics with $\xi_0 = 160~\mu$m. (a) Representative experimental OAM spectrum, after 20 cm of propagation for $|\delta_0| = 0.7$.  (b) Plot of $\Delta_+ - \Delta_-$ vs. $z$, where $\Delta_\pm = |A_{\mathrm{HP},\pm}|^2 - |A_{\mathrm{LP},\pm}|^2$. HP (LP) denotes $P_0 = 0.86$ W ($P_0 = 0.03$ W).  Dashed black line: numerics for $|\delta_0| = 0.7$. Solid blue line: numerics for $|\delta_0| = 1.4$. Black squares: experiment with  $|\delta_0| = 0.7$.  Blue diamonds: experiment with $|\delta_0| = 1.4$.  (c) Plot of $|\delta|$ vs. $\langle r \rangle$ from numerics with $|\delta_0| = 0.7$.  }
     \label{fig:OAMexp}
\end{figure}

\indent An interesting question arises in relation to whether an excited state with non-zero OAM ($|\delta| \neq 1$) can explore the full energy landscape (varying both $|\delta|$ and $\xi$)? Evolution of $|\delta|$ implies a change in the net angular momentum of the photon droplet. To reconcile this, the photon droplet is expected to excite higher-order OAM modes as it relaxes towards $|\delta| = 1$. As the system is (mostly) closed however, such excitations cannot escape and we expect that higher-order OAM excitations are re-absorbed by the droplet at a later ``time", leading to beating dynamics in both $\xi$ and $|\delta|$. To study this, we choose an initial state with $|\delta_0| = 0.7$ and $\xi_0 = 160~\mu$m, shown in Fig.~\ref{fig:rotation}.  Figure~\ref{fig:rotation}(a) shows experimental evolution of a beam with $P_0 = 0.03$ W to provide a linear propagation reference. Figures~\ref{fig:rotation}(b) and ~\ref{fig:rotation}(c) show the experimental and corresponding numerical profiles, respectively, for $P_0 = 0.86$ W.  Figure~\ref{fig:rotation}(d) shows plots of droplet rotation angle $\theta$ as a function of $z$ calculated from numerics using the input parameters of Figs.~\ref{fig:rotation}(a) and \ref{fig:rotation}(b). \\
\indent We demonstrate the dynamical evolution of the photon droplet by measuring the OAM spectrum of the beam at planes P0-P4.  We replace the camera with a second SLM (see Fig.~\ref{fig:exp}, inset), and perform an OAM mode decomposition \cite{Mair2001, Leach2010, Liu2015}.  We display a series of forked diffraction gratings onto the SLM associated with a given OAM number $\ell_{G}$, couple the first diffraction order into a single-mode fiber, and measure the intensity of the coupled light with a photodiode.  Only the component of the field without OAM ($\ell = 0$) efficiently couples into the single-mode fiber, such that for a diffraction grating with OAM $\ell_{G}$, the photo-diode measures $|A_{-\ell_G}|^2$, the amplitude of the component of $E(r,z)$ with OAM $\ell = -\ell_G$.  We also calculate $|A_\ell|^2 = |\int E e^{-i\ell\phi} r dr d\phi|^2$ from the corresponding numerical simulation, and normalize to $\sum |A_\ell|^2 = 1$.\\
\indent Figure~\ref{fig:OAMexp}(a) shows a representative experimental OAM spectrum $|A_\ell|^2$ versus OAM number $\ell$ after 20 cm of propagation, showing that the photon droplet loses OAM by coupling to higher-order OAM modes.  We quantify the net OAM in the photon droplet by calculating $\Delta_+ - \Delta_-$, where $\Delta_\pm = |A_{HP,\pm}|^2 - |A_{\mathrm{LP},\pm}|^2$, $+$ ($-$) denotes $\ell = +1$ ($\ell = -1$), HP denotes $P_0 = 0.86$ W, and LP denotes the linear reference with $P_0 = 0.03$ W.    Calculating $\Delta_+ - \Delta_-$ accounts for systematic error due to any small changes in alignment in the experimental OAM measurement.   Figure~\ref{fig:OAMexp}(b) shows plots of $\Delta_+ - \Delta_-$ vs. $z$ for $|\delta_0| = 0.7$ (numerics: dashed black line; data: black squares) and $|\delta_0| = 1.4$ (numerics: solid blue line; data: open blue diamonds) {\footnote{Experimental error for $\Delta_+ - \Delta_-$ is calculated by taking the standard deviation of the residuals from the fit to each measured amplitude $|A_{HP(LP), \pm}|^2$, then summing the errors in quadrature.}.  As seen in Fig.~\ref{fig:OAMexp}(b),  $\Delta_+ - \Delta_-$ varies with $z$, a clear demonstration that the relative balance of $|A_+|^2$ and $|A_-|^2$ changes as the photon droplet evolves.    The dynamical evolution of $|\delta|$ vs. $\langle r \rangle$ calculated from the complementary numerics and plotted in Fig.~\ref{fig:OAMexp}(c) shows how the relevant physical parameters evolve as the photon droplet explores its pseudo-energy landscape.

\indent In conclusion, we identify a robust self-bound photon droplet propagating in a nonlocal nonlinear medium. We show experimentally and numerically that a photon droplet displaced from equilibrium will exchange OAM back and forth with higher-order OAM modes as the photon droplet evolves in a manner reminiscent of the two-way energy exchange between a hot spot and the photon bath in optical filamentation \cite{Couairon2007}. The photon droplet description presented here provides a framework for exploring both a stationary soliton-like ground state, and the dynamics associated with higher-energy excited states. These classical photon droplet states share a remarkable similarity to quantum droplets of dipolar BECs and 1D liquid Helium, and may give insight into the behavior of quantum \cite{Gomez2014} and classical \cite{Hill2008} rotating droplets.
\begin{acknowledgments}
D.F. acknowledges financial  support from the European Research Council under the European
Union's Seventh Framework Programme (FP/2007-2013)/ERC GA 306559, the
Engineering and Physical Sciences Research Council (EPSRC, UK, grants EP/M006514/1,
EP/M01326X/1).  
N.W. and C.W.D. acknowledge support from EPSRC CM-CDT Grant No. EP/L015110/1.  M.V. and P. \"O. acknowledge support from EPSRC EP/M024636/1.
\end{acknowledgments}


%

%

\newpage
\newpage
\begin{center}
\textbf{\large Supplemental Materials}
\end{center}
\setcounter{equation}{0}
\setcounter{figure}{0}
\setcounter{table}{0}
\setcounter{page}{1}
\makeatletter
\renewcommand\thefigure{S\arabic{figure}}    

  \begin{figure}[h!]
\centering
\includegraphics[width=0.95\columnwidth]{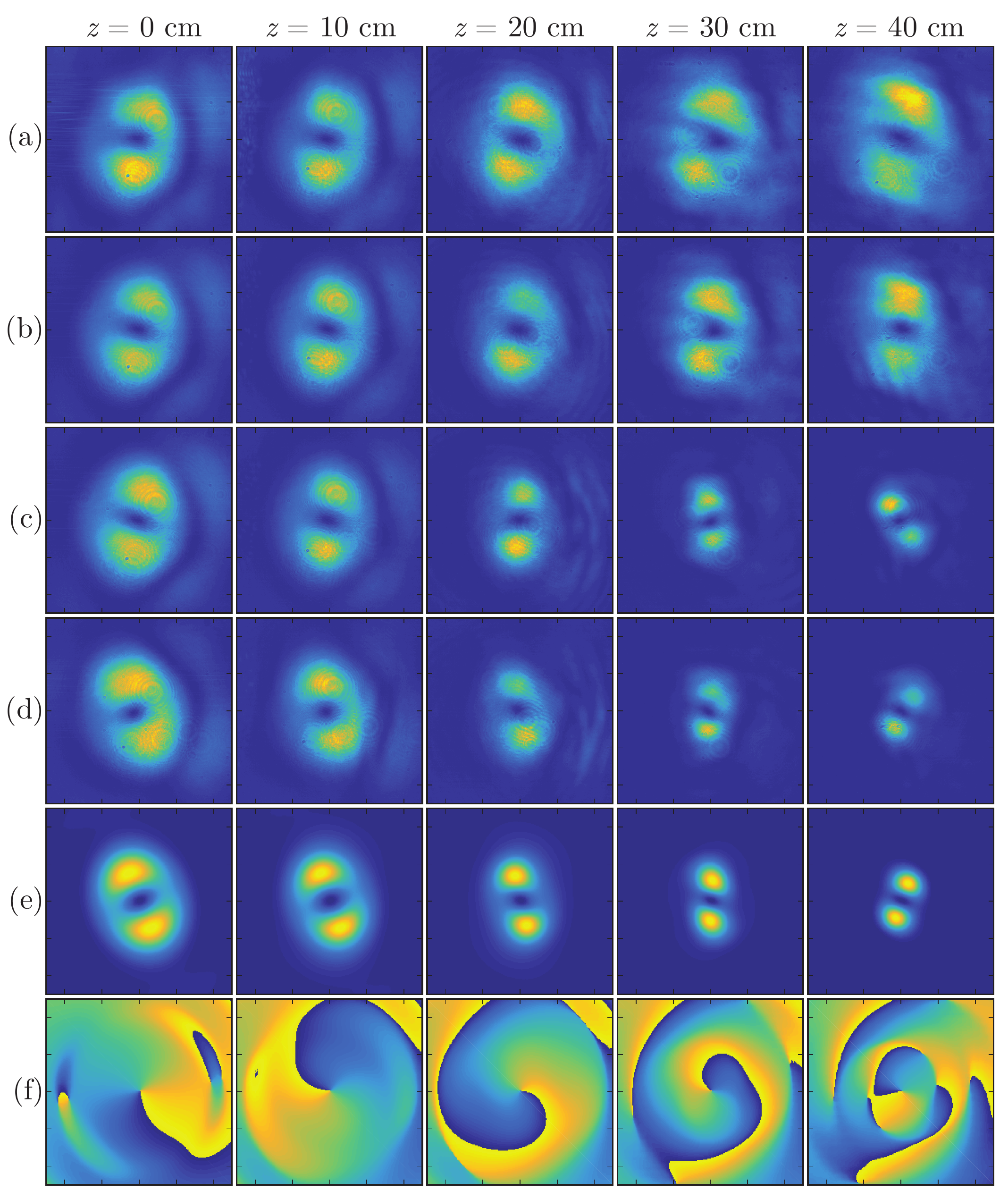}
 \caption{1-mm x 1-mm square images of the beam intensity profile as it propagates for varying input beam powers $P_0$.  (a) Linear beam propagation with $P_0 = 0.03$ W, $|\delta_0| = 0.2$, and $\xi_0 = 220~\mu$m. (b) Photon droplet with $P_0 = 0.36$ W, $|\delta_0| = 0.2$, and $\xi_0 = 220~\mu$m.  (c) Photon droplet with $P_0 = 0.86$ W, $|\delta_0| = 0.2$, and $\xi_0 = 220~\mu$m.  (d) Photon droplet with $P_0 = 0.86$ W, $|\delta_0| = 5$, and $\xi_0 = 220~\mu$m. The direction of droplet rotation switches for $|\delta_0| > 1$. (a)-(d) are from experiment. (e)-(f) Numerical simulation corresponding to experimental parameters for (d).   }
 \label{fig:images}
 \end{figure}

 The spatial evolution of the photon droplets corresponding to the curves in Fig.~\ref{fig:OAM} in the main text is shown and discussed below.  We experimentally explore the photon droplet evolution within the energy landscape as function of particle number.  The experimental images shown in Fig.~\ref{fig:images}(a)-(c) correspond to the initial conditions used for the numerics of Fig.~\ref{fig:OAM} in the main text, $|\delta_0| = 0.2$ and $\xi_0 = 220~\mu$m and show how the spatial profile of the photon droplet evolves for varying particle number (beam power $P_0$).  For $P_0 = 0.03$ W, as shown in Fig.~\ref{fig:images}(a) we observe expansion consistent with linear diffraction.  For $P_0 = 0.36$ W shown in Fig.~\ref{fig:images}(b) we observe slow droplet rotation and no focusing dynamics, consistent with the discussion in the main text for $\xi_0 \sim \xi_\mathrm{min}$ and relatively small nonlinearity.  At $P_0 = 0.86$ W, the photon droplets shown in Fig.~\ref{fig:images}(c)-(f) start in a highly excited state both in terms of $\xi_0$ and $|\delta_0|$, such that $|\delta|$ and $\langle r \rangle \propto \xi$ vary simultaneously. 
  
We also observe rotation of the photon droplets, most evident for the highly excited droplets shown in Fig.~\ref{fig:images}(d)-(f).  The direction of the rotation is dependent on the net photon droplet OAM as shown in Figs~\ref{fig:images}(c) and \ref{fig:images}(d), corresponding to $|\delta_0| = 0.2$ and $|\delta_0| = 5$ respectively.   Figures~\ref{fig:images}(e) and \ref{fig:images}(f) show the intensity (e) and phase (f) profiles from numerical simulations  corresponding to the experimental images shown in  Fig.~\ref{fig:images}(d).   Figure~\ref{fig:rotation} shows plots of rotation angle $\theta$ versus propagation distance $z$ for varying input beam powers $P_0$.  Plots labeled (a),(b),(c), and (d-f) are calculated from numerical simulations and correspond to the same input parameters as the experimental beam profiles shown in Fig.~\ref{fig:images}.  We note that even for the linear case there is a small amount of residual rotation.  This is due to having some small extraneous population of higher order OAM modes due to the way that we form our input states.

 \begin{figure}[h!]
\centering
\includegraphics[width=0.95\columnwidth]{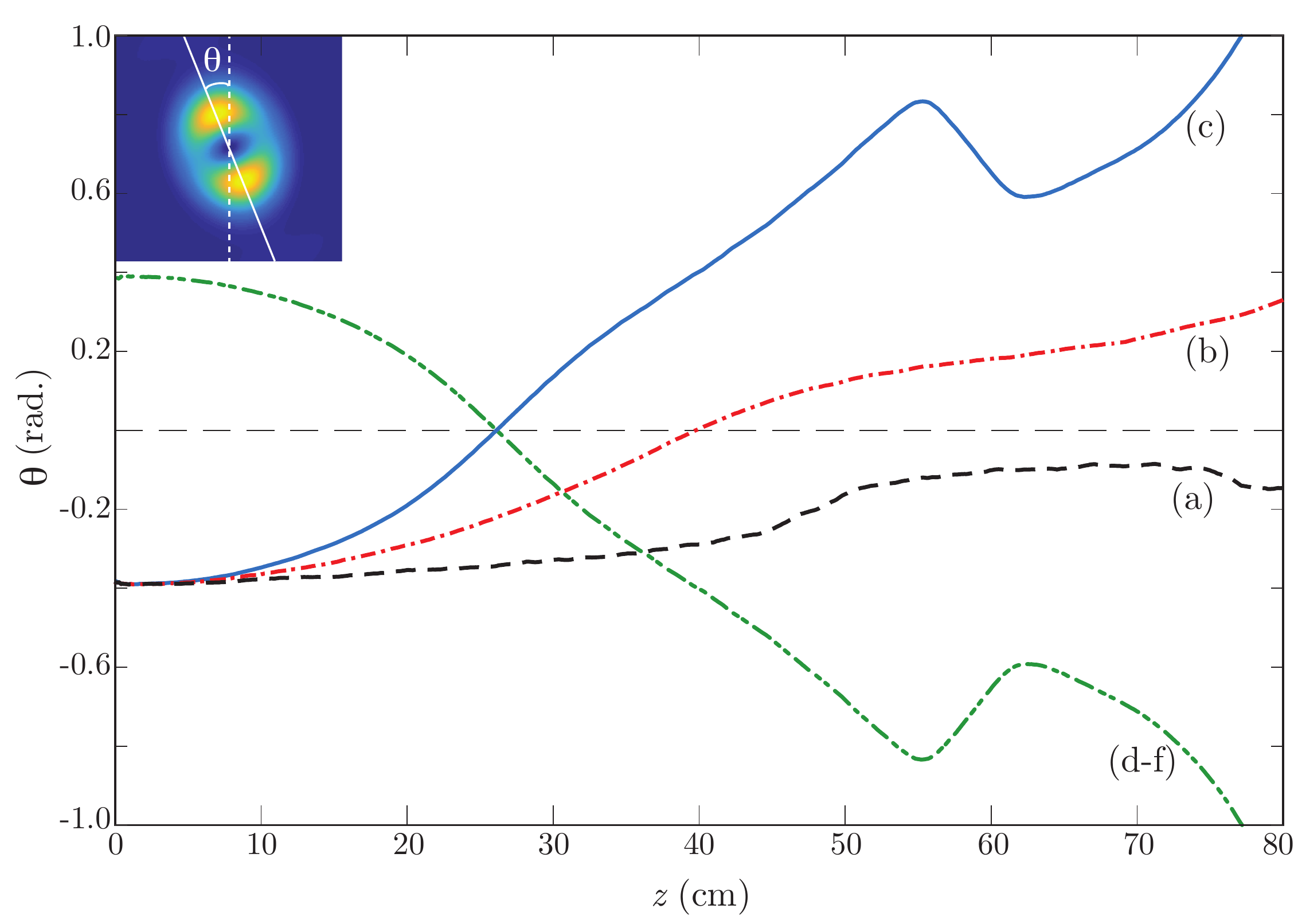}
 \caption{Plots of rotation angle $\theta$ versus $z$ for varying input beam powers $P_0$, calculated from numerical simulations.   (a) Linear beam propagation with $P_0 = 0.03$ W, $|\delta_0| = 0.2$, and $\xi_0 = 220~\mu$m. (b) Photon droplet with $P_0 = 0.36$ W, $|\delta_0| = 0.2$, and $\xi_0 = 220~\mu$m.  (c) Photon droplet with $P_0 = 0.86$ W, $|\delta_0| = 0.2$, and $\xi_0 = 220~\mu$m.  (d) Photon droplet with $P_0 = 0.86$ W, $|\delta_0| = 5$, and $\xi_0 = 220~\mu$m. The direction of droplet rotation switches for $|\delta_0| > 1$.  }
 \label{fig:rotation}
 \end{figure}

\end{document}